\documentstyle[12pt]{article}
\newcommand{\nn}{\nonumber\\}\newcommand{\p}[1]{(\ref{#1})}
\def\PRL#1#2#3{{\sl Phys. Rev. Lett.} {\bf#1} (#2) #3}
\def\NPB#1#2#3{{\sl Nucl. Phys.} {\bf B#1} (#2) #3}

\def\PRD#1#2#3{{\sl Phys. Rev.} {\bf D#1} (#2) #3}

\def\PLB#1#2#3{{\sl Phys. Lett.} {\bf #1B} (#2) #3}

\def\LMP#1#2#3{{\sl Letters in Math. Phys.} {\bf #1} (#2) #3}

\def\TMP#1#2#3{{\sl Theor. Math. Phys.} {\bf #1} (#2) #3}

\def\JETPL#1#2#3{{\sl  Sov. Phys. JETP Lett.} {\bf #1} (#2) #3}

\def\CQG#1#2#3{{\sl Class. Quantum Grav.} {\bf #1} (#2) #3}

\def\Ai{\hbox{\hbox{${\cal A}$}}\kern-1.9mm{\hbox{${/}$}}}
\def\Vi{\hbox{\hbox{${\cal V}$}}\kern-1.9mm{\hbox{${/}$}}}
\def\Di{\hbox{\hbox{${\cal D}$}}\kern-1.9mm{\hbox{${/}$}}}
\def\lam{\hbox{\hbox{${\lambda}$}}\kern-1.6mm{\hbox{${/}$}}}
\def\D{\hbox{\hbox{${D}$}}\kern-1.9mm{\hbox{${/}$}}}
\def\A{\hbox{\hbox{${A}$}}\kern-1.8mm{\hbox{${/}$}}}
\def\V{\hbox{\hbox{${V}$}}\kern-1.9mm{\hbox{${/}$}}}
\def\parz{\hbox{\hbox{${\partial}$}}\kern-1.7mm{\hbox{${/}$}}}
\def\B{\hbox{\hbox{${B}$}}\kern-1.7mm{\hbox{${/}$}}}
\def\R{\hbox{\hbox{${R}$}}\kern-1.7mm{\hbox{${/}$}}}
\def\si{\hbox{\hbox{${\xi}$}}\kern-1.7mm{\hbox{${/}$}}}

\begin{document}
\thispagestyle{empty}
\renewcommand{\thefootnote}{\dagger}

\centerline
{\bf DOUBLY SUPERSYMMETRIC NULL STRINGS AND}
\centerline
{\bf STRING TENSION GENERATION}

\vskip 1truecm

\centerline
{\bf Igor A. Bandos}

\bigskip

\centerline
{\sl Kharkov Institute of Physics and Technology}
\centerline
{\sl Kharkov, 310108, the Ukraine}

\vskip 0.5truecm

\centerline
{\bf Dmitrij P. Sorokin, \footnote{Supported in part by the European Community
Research Program ``Gauge Theories, Applied Supersymmetry and Quantum
Gravity'' under contract CEE-SCI-CT92-0789}
\renewcommand{\thefootnote}{\ddagger}
\footnote{Permanent address: Kharkov Institute of Physics and
Technology, Kharkov, 310108, the Ukraine
e-mail address:
kfti\%kfti.kharkov.ua@relay.ussr.eu.net}
\hskip12pt Mario Tonin}

\bigskip

\centerline{\sl Dipartimento di Fisica ``G. Galilei"}
\centerline{\sl Universit\`a  degli Studi di Padova, Italy}

\vskip 0.5truecm

\centerline
{\bf Dmitrij V. Volkov}

\bigskip

\centerline
{\sl Kharkov Institute of Physics and Technology}
\centerline
{\sl Kharkov, 310108, the Ukraine}

\vskip 1.5truecm

\noindent
DFPD/93/TH/48\\
July 1993

\vskip 1.5truecm

\noindent
{\bf Abstract.} We propose a twistor--like formulation of N=1,
D=3,4,6 and 10 null superstrings. The model possesses N=1 target
space supersymmetry and n=D--2 local worldsheet supersymmetry,
the latter replaces the $\kappa$--symmetry of the conventional approach
to the strings. Adding a Wess--Zumino term to a null superstring
action we observe a string tension generation mechanism \cite{t,dg}:
the induced worldsheet metric becomes non--degenerate and the
resulting model turns out to be classically equivalent to the heterotic
string.

\setcounter{page}1
\renewcommand{\thefootnote}{\arabic{footnote}}
\setcounter{footnote}0
\newpage

Null strings \cite{shi} have attracted certain attention from various
points of view  \cite{zh1,zh2,noc,zb,bn,l,t}.
One may consider them as a toy model for studying
ordinary strings since the former, being characterized by the zero tension
and degenerate worldsheet metric, take an intermediate position between
the particles and the strings. Actually, the null strings are closer to
the particles than to the strings because they just describe a collection
of particles moving on a null surface  \cite{zh1,l}.
Moreover the null strings do
not need any critical dimension of space--time for living in
\cite{noc,zb} and
the null superstrings  \cite{zh1,bn,l}, in contrast to the ordinary
superstrings, do
not require a Wess--Zumino term for getting the fermionic $\kappa$--symmetry,
which is necessary to get rid of redundant fermionic degrees of freedom
(see \cite{gsw} for a review on strings).

{}From a physical point of view the most interesting fact is an assumption
that the null string may be regarded as a high energy limit of the string
 \cite{zh2,l}, thus providing a way for describing strings beyond the Planck's
scale.

Here, performing a program of ``twistorizing everything''  \cite{stv}--
\cite{cp}, \cite{dg} we
propose a twistor--like formulation of N=1, D=3,4,6 and 10 null
superstrings. The model possesses N=1 target space supersymmetry and
n=D--2 local worldsheet supersymmetry, the latter replaces the
$\kappa$--symmetry of other approaches
\footnote{Earlier various doubly
supersymmetric models were considered in  \cite{double} irrelevant to the
$\kappa$--symmetry problem.}.

We shall see that for the doubly super\-sym\-metric null string ac\-tion to be
super re\-pa\-ra\-met\-ri\-za\-tion inva\-riant one needs only a
su\-per\-field containing
a bosonic ``einbein". This einbein characterizes the geometry of the
bosonic part of the null super worldsheet  \cite{zh1,zb,l}.
So the worldsheet gravitino
and O(n) gauge field drop out of the null string action.

The geometry of the null super worldsheet \cite{l,lr} resembles that of
a spinning particle
worldline  \cite{b}  from the one--hand side and the heterotic worldsheet
geometry  \cite{tonin,dg} from the other. The action of the model under
consideration
turns out to be closely related to a ``geometro--dynamical" part (in
terminology of Galperin and Sokatchev  \cite{gs2}) of a heterotic string
twistor--like action  \cite{tonin,dg}.
The former arises as a singular solution of
twistor--like heterotic string constraints which corresponds to a
degenerate worldsheet metric and requires the string tension to be
zero\footnote{For N=2, D=3 Green--Schwarz superstring a similar singular
solution was considered in  \cite{gs2}. But since the authors imposed too
rigorous gauge fixing they associated the solution  with a superparticle
and not a null superstring.}.

On the contrary, adding a Wess--Zumino term generating the string tension
to the null superstring action is similar to a tension generation
mechanism proposed in  \cite{t,dg}. Note that earlier an alternative way of
generating the string tension was considered in  \cite{zh2}.

Our starting point is a twistor--like action for a bosonic null string
written in a first--order formalism:
\begin{equation}\label{1}
S=\int d^2\xi p_m(V^\mu\partial_\mu x^m-\bar\lambda \gamma^m\lambda),
\end{equation}
where $\xi^\mu (\mu=0,1)$ parametrize a null surface whose geometry is
determined by an ``einbein" $V^\mu$ (see \cite{l} for the details),
$x^m$ and $p_m$ are, respectively, the null string coordinate and the
momentum density in D=3,4,6 or 10 dimensional target space--time
(m=0,1,...,D--1), and $\lambda^\alpha$ is a Majorana or
Majorana--Weyl commuting spinor (twistor component) depending on the
dimension of the target space.

Action \p{1} is a generalization of the twistor--like particle action
 \cite{stv,gs}, and can be derived from other formulations of the null
string  \cite{zh1,zb,l}.

To show that eq.~\p{1} indeed describes a bosonic null string we solve
the equations of motion:
$$
\partial_\mu (V^\mu p_m) = 0,
$$
\begin{equation}\label{2}
V^\mu\partial_\mu x^m - \bar\lambda \gamma^m \lambda=0,
\end{equation}
\begin{equation}\label{3}
p_m(\bar\lambda \gamma^m)_\alpha=0\rightarrow p_m = a
(\xi) \bar\lambda\gamma_m \lambda,
\end{equation}
\begin{equation}\label{4}
p_m \partial_\mu x^m = 0,
\end{equation}
where \p{3} represents the Cartan relation for the lightlike vector
in D=3,4,6 and 10. From eqs.~\p{2}--\p{4} it follows that
\begin{equation}\label{5}
V^\mu \partial_\mu x^m \partial_\nu x_m = 0.
\end{equation}

This means that the induced worldsheet metric $G_{\mu\nu}=\partial_\mu
x^m \partial_\nu x_m$ is degenerate, which is a peculiar property of the
null string.

If we introduce intrinsic degenerate metric $g_{\mu\nu}$ on the
null string
\begin{equation}\label{5.a}
V^\mu g_{\mu\nu}=0,
\end{equation}
eqs.~\p{5} and \p{5.a} mean that $G_{\mu\nu}$ and $g_{\mu\nu}$ coincide up to
a scalar factor. The degeneracy of the metric implies that there exists
a vector $\Omega_\mu$ such that
\begin{equation}\label{5.b}
V^\mu \Omega_\mu = 0 \quad {\rm and}\quad
g_{\mu\nu} = \Omega_\mu \Omega_\nu.
\end{equation}

One may also assume that on the worldsheet there live vectors $V_\mu$ and
$\Omega^\mu$ inverse to $V^\mu$ and $\Omega_\mu$, respectively:
\begin{equation}\label{5.c}
V^\mu V_\mu = 1 = \Omega_\mu \Omega^\mu
\end{equation}
We use all these vectors to construct a tension generating term in
the action of null superstring. Note that a priori we do not require
these vectors to form a non-degenerate  intrinsic metric on the
worldsheet.
Now let us show how a tension generating mechanism works
for the simple case of the bosonic null string. Following
Refs.~\cite{t,dg}
we introduce an ``electromagnetic" field $A_\mu$ on a null surface and
put it into the action \p{1} as part of the following term
\begin{equation}\label{5.d}
S=\int d^2\xi [ p_m (V^\mu\partial_\mu x^m - \bar\lambda \gamma^m\lambda)+
\phi\epsilon^{\mu\nu}\left(\partial_\mu A_\nu - V_\mu\Omega_\nu
(V^\lambda\partial_\lambda x^m)(\Omega^\rho\partial_\rho x_m)\right)]
\end{equation}
where $\phi(\xi)$ is a Lagrange multiplier and $\epsilon^{12} = -
\epsilon^{21} =1$. Varying \p{5.d} with respect to $\phi$ we just find that
the strength of $A_\mu$ is:
\begin{equation}\label{5.e}
F\equiv\epsilon^{\mu\nu} \partial_\mu A_\nu = \epsilon^{\mu\nu}
V_\mu \Omega_\nu (V^\lambda\partial_\lambda x^m)
(\Omega^\rho \partial_\rho x_m).
\end{equation}
Varying with respect to $A_\mu$ gives
\begin{equation}\label{5.f}
\partial_\mu\phi=0 \rightarrow \phi = T,
\end{equation}
where $T$ is a constant identified with the string tension \cite{t,dg}.
Substituting \p{5.f} into the action \p{5.d} and skipping the total
derivative term we get:
\begin{equation}\label{5.g}
S=\int d^2\xi [ p_m (V^\mu\partial_\mu x^m - \bar\lambda \gamma^m
\lambda) - T\epsilon^{\mu\nu} V_\mu\Omega_\nu (V^\lambda
\partial_\lambda
x^m)(\Omega^\rho\partial_\rho x_m)]
\end{equation}
Observe that eq.~\p{5.d} acquires additional symmetry found in
\cite{tonin}:
\begin{equation}\label{5.i}
\delta p_m  =\phi\epsilon^{\mu\nu}V_\mu\Omega_\nu
A(\xi)(V^\mu\partial_\mu x^m + \bar\lambda\gamma^m
\lambda)\qquad
\delta\Omega^\rho = - A(\xi) V^\rho,\qquad \delta V^\mu=0.
\end{equation}
This symmetry allows  one to eliminate $p_m$ on the mass shell \p{2},
\p{3}. With this in mined,
from the variation of \p{5.g} with respect to $V^\mu$ (and  taking into
account \p{5.c}) one gets the second Virasoro condition (the first
one follows from \p{2}):
\begin{equation}\label{5.j}
\Omega^\mu \partial_\mu x_m \Omega^\rho \partial_\rho x^m =0.
\end{equation}

As a result the induced metric $G_{\mu\nu}=\partial_\mu x^m \partial_\nu
x_m$ is not degenerate anymore, and the null string is transformed into
an ordinary bosonic string, with $V_\mu=e^+_\mu, \Omega_\mu=e^-_\mu$
being identified with the nondegenerate worldsheet zweibeins. Upon
dropping the $p_m$ term out of \p{5.g} one restores the conventional bosonic
string action.

Let us proceed now with null superstrings.
To get an action for a null superstring we extend the null worldsheet
geometry to a geometry on a null super worldsheet parametrized by $\xi^\mu$
and by n=D--2 odd variables $\eta^a(a=1,...,n)$, and consider $p_m,
x^m, V^\mu$ and $\lambda^\alpha$ as components of worldsheet superfields
to be introduced below. Note that on the null d=2 surface the spinors are
actually scalars, so the letters $a,b,c,..$ from the beginning of the
Latin alphabet correspond to the local $O(n)$ group indices \cite{l,lr}.

In contrast to Ref.~\cite{lr}, where a number of null d=2 superspaces
were studied in
detail, our choice of constraints on the super worldsheet vielbeins is
unique for all n--extended supersymmetries and reads as follows
\begin{equation}\label{6}
\{ \nabla_a, \nabla_b\} - (E^{-1})^d_c (\nabla_d E^c_{\{a})
\nabla_{b\}} = \delta_{ab}\Delta
\end{equation}
where $\nabla_a = E_a^b \partial_b + E_a^\mu \partial_\mu$,
$\Delta=E^\mu \partial_\mu + E^b \partial_b$, are the supercovariant
spinor and vector derivative, respectively, and $E_a^b, E^b, E_a^\mu,
E^\mu$ are the components of the super worldsheet vielbeins; $\{...\}$
denotes symmetrization. The second
term in the left--hand side of \p{6} corresponds to an $O(n)$ connection in
the tangent space to the null super worldsheet. Eq.~\p{6} allows one to
express $E^a$ and $E^\mu$ in terms of $E^b_a$ and $E^\mu_a$.

Redefining $E^\mu_a$ as
\begin{equation}\label{7}
E^\mu_a\equiv i E^b_a V^\mu_b
\end{equation}
and using superdiffeomorphisms and local $O(n)$ transformations we can
restrict the symmetry transformations on the null super worldsheet to
supergravity transformations by choosing a gauge analogous to that used
in the case of spinning particles \cite{b}:
\begin{equation}\label{8}
E^b_a = \delta^b_a E(\xi,\eta),
\end{equation}
where $E(\xi,\eta)$ is a superfield containing a gravitino
field $\chi_i(\xi)$ and $O(n)$ gauge field $A_{[ij]}(\xi)$ as
components. (In the null superspace $\chi_i$ does not carry vector
indices \cite{l,lr}.

With eqs.~\p{7} and \p{8} the constraints \p{6} are
reduced to
\begin{equation}\label{9}
D_a V^\mu_b + D_b V^\mu_a = {2\over n} \delta_{ab} V^\mu_c,
\end{equation}
where
\begin{equation}\label{10}
D_a=\partial_a + i V_a^\mu \partial_\mu,\quad
\{ D_a, D_b\} = {2i\over n} \delta_{ab}
(D_c V^\mu_c)\partial_\mu.
\end{equation}

The local supersymmetry transformations look as follows
\begin{eqnarray}\label{11}
\delta \eta_a = - {i\over 2} D_a \Lambda (\xi,\eta),\qquad
\delta\xi^\mu = \Lambda^\mu(\xi,\eta) - {1\over 2} V^\mu_a
D_a\Lambda\nn
\delta V^\mu_a  = -i D_a \Lambda^\mu + {i\over 2} (D_b\Lambda)
(D_a V^\mu_b),\qquad \delta D_a = {i\over 2} (D_a D_b \Lambda) D_b
\end{eqnarray}
where $\Lambda(\xi,\eta)$ and $\Lambda^\mu(\xi,\eta)$ are superfield
parameters.

In a Wess--Zumino gauge
\begin{equation}\label{13}
D_a |_{\eta=0} = \partial_a
\end{equation}
with the use of local sym\-metry pa\-ra\-meters
$$D_{[a_1} D_{b_1]}
\Lambda^\mu|_{\eta=0},..., D_{[a_1}D_{a_2}...D_{a_{n-1}]}\Lambda^\mu
|_{\eta=0}
$$
one may put all components of $V^\mu_a$ to zero except
\begin{equation}\label{14}
D_a V^\mu_a|_{\eta=0} = V^\mu,
\end{equation}
where $V^\mu$ is the null worldsheet einbein (eq.~\p{1}). Note that
$V^\mu_a$ is a generalization of an $E^+_\alpha$ superfield used
in \cite{dg} for defining heterotic geometry.

Now we are ready to write down a doubly supersymmetric generalization
of the null string action \p{1}.\footnote{For simplicity we consider
flat geometry in the target superspace, though the generalization to the
case of a supergravity background is straightforward}
 It looks very much like a twistor--like
superparticle action \cite{stv,gs} and/or heterotic string action
\cite{tonin,dg} without a Wess-Zumino part:
\begin{equation}\label{15}
S_N=\int d^2\xi d^n\eta {\rm P}_{am}
(D_a X^{m} -i D_a \bar\Theta\gamma^{m} \Theta).
\end{equation}

In \p{15} a bosonic superfield $X^m (X^m|_{\eta=0}=x^m)$ and a
fer\-mi\-onic su\-per\-fi\-eld
$\Theta^\alpha(\Theta^\alpha|_{\eta=0} =
\theta^\alpha$, $D_a\Theta^\alpha|_{\eta=0}=\lambda^\alpha_a)$
transform as scalars under the local supersymmetry transformations
\p{11}, and the Lagrange multiplier superfield ${\rm P}_{am}$ transforms in
such a way that the action \p{15} is invariant under the local supersymmetry
transformations.

Observe that the superfield \p{8} dropped out of the action, thus the
latter does not contain the gravitino and the $O(n)$ gauge field. It
is amusing that space-time supersymmetry naturally arises as a consequence
of the worldsheet supersymmetrization of eq.~\p{1}. Thus, once again
worldsheet geometry proves itself to be more fundamental than the
target space one. So eq.~\p{15} is also invariant under the global
supersymmetry transformations in the target space
\begin{equation}\label{16}
\delta\Theta^\alpha= \varepsilon^\alpha,\qquad \delta X^m = \bar\Theta\gamma^m
\varepsilon,
\end{equation}
and local transformations of the Lagrange multiplier
\begin{equation}\label{17}
\delta {\rm P}_{am}=\left(D_b+i\partial_\mu
V^\mu_b\right)\bar\Sigma_{abc}\gamma^mD_c\Theta,
\end{equation}
where $\Sigma_{abc}^\alpha$ is totally symmetric and traceless with
respect to the $O(n)$ indices.

Following the reasoning course performed in detail for superparticles
\cite{gs}
and heterotic strings \cite{tonin,dg} one may derive that all higher
components of
$X^m$ and $\Theta^\alpha$ are expressed in terms of leading components,
and the only component of ${\rm P}_{am}$ which survives gauge fixing
\p{17} and is not equal to zero due to equations of motion
$$
\left(D_a+i\partial_\mu V^\mu_a\right){\rm P}_{am}=0,
$$
\begin{equation}\label{18}
\left({\rm P}_{am}D_a\bar\Theta\gamma^m\right)_\alpha=0
\end{equation}
is the null string momentum density
\begin{equation}\label{19}
p_m=\varepsilon_{aa_{1}...a_{n-1}}\left(D_{a_{1}}+i\partial_\mu
V^\mu_{a_{1}}\right)...\left(D_{a_{n-1}}+i\partial_\mu V^\mu_{a_{n-1}}\right)
{\rm P_{am}}\vert_{\eta=0}.
\end{equation}
Thus, in the Wess-Zumino gauge \p{13}, \p{14} the action \p{15} is reduced to
the following component form:
\begin{equation}\label{20}
S=\int d^2\xi p_m\left(V^\mu\left(\partial_\mu x
^m-i\partial_\mu\bar\theta\gamma^m\theta\right)-{1\over
n}\bar\lambda_a\gamma^m\lambda_a\right),
\end{equation}
where $\lambda_a^\alpha$ are subject to the constraint \cite{gs}
\begin{equation}\label{21}
\bar\lambda_a\gamma^m\lambda_b={1\over n}\delta_{ab}
\bar\lambda_c\gamma^m\lambda_c.
\end{equation}

Then eliminating the $\lambda$--term by use of eq.~\p{3}, integrating
over $p_m$ and rescaling $V^\mu$ we get a form of the null superstring
which one may find, for
example, in Refs.~\cite{l}:
\begin{equation}\label{22}
S=\int d^2\xi V^\mu V^\nu\left(\partial_\mu x
^m-i\partial_\mu\bar\theta\gamma^m\theta\right)\left(\partial_\nu x
^m-i\partial_\nu\bar\theta\gamma^m\theta\right).
\end{equation}

The action \p{22} is invariant under $\kappa$--symmetry transformations
\cite{zh2,l}:
$$
\delta_\kappa\theta^\alpha=iV^\mu\left(\partial_\mu x
^m-i\partial_\mu\bar\theta\gamma^m\theta\right)
\gamma_{m\beta}^\alpha\kappa^\beta,\hskip24pt \delta_\kappa
x^m=i\bar\theta\gamma^m\delta_\kappa\theta,
$$
\begin{equation}\label{23}
\delta_\kappa V^\mu=2V^\mu V^\nu\partial_\nu\bar\theta\kappa.
\end{equation}

Using the twistor condition $V^\mu\left(\partial_\mu x
^m-i\partial_\mu\bar\theta\gamma^m\theta\right)=
{1\over n}\bar\lambda_a\gamma^m\lambda_a$ we may replace the parameter
$\kappa^\beta$ by a parameter
$D_a\Lambda\vert_{\eta=0}\equiv\epsilon_a=\bar\lambda_a\kappa$ so that, in
particular,
\begin{equation}\label{24}
\delta\theta^\alpha=\epsilon_a\lambda^\alpha_a;\hskip24pt \delta x^m=
i(\bar\theta\gamma^m\lambda_a)\epsilon_a.
\end{equation}
In eqs.~\p{24} one may recognize the on-shell $n=D-2$ local supersymmetry
transformations \p{11} of the $\theta^\alpha$ and $x^m$ components, which
allow one to eliminate half of the fermionic degrees of freedom of the
null superstring.
Thus, $n=D-2$ extended local supersymmetry of the null super worldsheet
takes upon itself the role of the $\kappa$--symmetry, and is in fact more
fundamental than the latter. This opens the
possibility of solving the problems connected with the infinite
irreducibility of the $\kappa$--symmetry constraints.

We have already mentioned above that the null superstring action \p{15}
may be obtained from the twistor-like heterotic string action
\cite{tonin,dg} by
taking an appropriate zero tension limit which results in vanishing of
the heterotic string Wess--Zumino term.

Let us now perform an inverse procedure by adding to eq.~\p{15} a
Wess--Zumino term generalizing that of the bosonic model \p{5.d} in such a
way that the heterotic string action is restored. This is achieved by the
string tension generating mechanism of Refs.~\cite{t,dg}.

The Wess--Zumino part of the action is written as follows:
\begin{equation}\label{25}
S_{WZ}=\int d^2\xi d^n\eta{\rm P}^{\cal {MN}}\left(\partial_{\cal M}A_{\cal
N}+B_{\cal{MN}}-F_{\cal {MN}}\right),
\end{equation}
where ${\cal {M,N}}$ stand for $d=2$ bosonic ($\mu,\nu,...)$ and $O(n)$
(a,b,...) fermionic indices of the worldsheet; ${\rm P}^{\cal {MN}}=(-1)^{{\cal
{MN}}+1}{\rm P}^{\cal {NM}}$ is a Lagrange multiplier; $A_{\cal M}$ is a
supersymmetric generalization of the "electromagnetic" field
$A_\mu$ in \p{5.b};
\begin{equation}\label{26}
B_{\cal {MN}}=i\partial_{\cal M}X^m\partial_{\cal
N}\bar\Theta\gamma_m\Theta+(-1)^{{\cal
{MN}}+1}i\partial_{\cal N}X^m\partial_{\cal M}\bar\Theta\gamma_m\Theta
\end{equation}
defines the pull-back onto the super worldsheet of the Wess--Zumino
two-form $B=dX^m\wedge d\bar\Theta\gamma_m\Theta$, which is the same as
that of the heterotic string \cite{dg}, and which, in general, can be
introduced in the case of a null superstring as well; and
\begin{equation}\label{27}
F_{\cal MN}=\left(V_{\cal M}\Omega_{\cal
N}+(-1)^{{\cal MN}+1}V_{\cal N}\Omega_{\cal M}\right)
D_a\bar\Theta\gamma_mD_a\Theta^{\cal L}\left(\partial_{\cal L}X^m-
i\partial_{\cal L}\bar\Theta\gamma_m\Theta\right)
\end{equation}
is introduced in such a way that, with taking into account the twistor
condition $D_aX^m-iD_a\bar\Theta\gamma_m\Theta=0$, its corollaries and
the constraint \p{9}, $dB=dF$ on the mass shell \cite{dg}.

In \p{27} superfields $V_{\cal M}$, $\Omega_{\cal M}$ and $\Omega^{\cal M}$
generalize the bosonic vectors \p{5.c}. In particular,
$$
V_\mu(D_aV^\mu_a)=1,\qquad
\Omega_\mu(D_aV^\mu_a)=0,\qquad
V_a=-iV^\mu_aV_\mu,\qquad \Omega_a=-iV^\mu_a\Omega_\mu.
$$

As in the bosonic case above one may convince oneself
that the induced worldsheet metric is not degenerate anymore, $V_{\cal
M}$ and $\Omega_{\cal M}$ may be identified with the elements
$E^+_{\cal M}$ and
$E^-_{\cal M}$ of the superzweibein matrix on the worldsheet, and the
action \p{15} plus \p{25} turns out to be completely the same as the
heterotic string action of Ref.~\cite{dg}. The string tension $T$ arises as an
integration constant of the equation of motion:
$$
\partial_{\cal M}{\rm P}^{\cal MN}=0,
$$
from which, by gauge fixing the local transformations $\delta{\rm P}^{\cal
MN}=\partial_{\cal L}\Xi^{\cal LMN}$ (where $\Xi^{\cal LMN}$
is a graded antisymmetric parameter), one may find that the only
nontrivial component of ${\rm P}^{\cal MN}$ is
$p^{\mu\nu}=\varepsilon^{\mu\nu}T\eta^n$ (see \cite{dg} for the details).

In conclusion we have constructed the doubly supersymmetric model  for
$N=1$, $D=3,4,6$ and 10 null superstrings which elucidates
from somewhat different point of view
 the meaning of the both ingredients of the twistor-like heterotic
string formulation \cite{tonin,dg}: the twistor-like null superstring action
corresponds to the ``geometro-dynamical'' part of
the twistor-like heterotic string action, while introducing an
"electromagnetic" superfield propagating in the null super worldsheet
and taking into account the Wess--Zumino term \'a la Townsend and
Galperin et. al. leads to the string tension generation and results in
the model which is proved to be classically equivalent to the heterotic
string.

\vspace{0.5cm}
{\large\bf Acknowledgments}

\medskip
D.S. is grateful to the European Community for financial support under
the contract CEE-SCI-CT92-0789, the University of Padova and INFN
(Sezione di Padova), and in particular Prof. M. Tonin for kind
hospitality in Padova.

I.B., D.S. and D.V. are grateful to A. Zheltukhin for stimulating
discussion.

\end{document}